\documentclass[9pt,twocolumn,twoside]{pnas-new}

\templatetype{pnasbriefreport} 

\title{Universal law for the vibrational density of states of liquids}

\author[a,b,c]{Alessio Zaccone}
\author[d,e,f]{Matteo Baggioli}

\affil[a]{Department of Physics "A. Pontremoli", University of Milan, via Celoria 16,
20133 Milan, Italy.}
\affil[b]{Department of Chemical Engineering and Biotechnology,
University of Cambridge, Philippa Fawcett Drive, CB30AS Cambridge, U.K.}
\affil[c]{Cavendish Laboratory, University of Cambridge, JJ Thomson
Avenue, CB30HE Cambridge, U.K.}

\affil[d]{Instituto de Fisica Teorica UAM/CSIC, c/Nicolas Cabrera 13-15,
Universidad Autonoma de Madrid, Cantoblanco, 28049 Madrid, Spain.
}
\affil[e]{Wilczek Quantum Center, School of Physics and Astronomy, Shanghai Jiao Tong University, Shanghai 200240, China}
\affil[f]{Shanghai Research Center for Quantum Sciences, Shanghai 201315.}

\leadauthor{Zaccone} 

\authorcontributions{A.Z.and M.B. equally contributed to the design and implementation of the research, to the analysis of the
results and to the writing of the manuscript.}
\authordeclaration{There are no competing interests to declare.}
\correspondingauthor{\textsuperscript{2}To whom correspondence should be addressed. E-mail: alessio.zaccone@unimi.it}


\begin{abstract}
An analytical derivation of the vibrational density of states (DOS) of liquids, and in particular of its characteristic linear in frequency low-energy regime, has always been elusive because of the presence of an infinite set of purely imaginary modes -- the instantaneous normal modes (INMs). By combining an analytic continuation of the Plemelj identity to the complex plane with the purely overdamped dynamics of the INMs, we derive a closed-form analytic expression for the low-frequency DOS of liquids. The obtained result explains from first principles the widely observed linear in frequency term of the DOS in liquids, whose slope appears to increase with the average lifetime of the INMs. The analytic results are robustly confirmed by fitting simulations data for Lennard-Jones liquids, and they also recover the Arrhenius law for the average relaxation time of the INMs, as expected.
\end{abstract}

\dates{This manuscript was compiled on \today}
\doi{\url{www.pnas.org/cgi/doi/10.1073/pnas.XXXXXXXXXX}}

\begin{document}

\maketitle
\thispagestyle{firststyle}
\ifthenelse{\boolean{shortarticle}}{\ifthenelse{\boolean{singlecolumn}}{\abscontentformatted}{\abscontent}}{}

In quantum physics, an exponentially decaying state is characterized by a complex value of the frequency, with the imaginary part giving the lifetime of the particle and its decay probability~\cite{Zeldovich}. Well known examples of unstable states with overwhelming imaginary part arise in the nuclear $\alpha$ decay (Gamow states) and, in particle physics -- the $W$ and $Z^0$ bosons~\cite{Stuart} -- where they are usually called \textit{resonances}. Moreover, within hydrodynamics and gravitational theories, these excitations are labelled \textit{quasinormal modes} \cite{Berti:2009kk} and they are the responsible for the black holes ringdown recently observed by Ligo and Virgo \cite{PhysRevLett.125.101102}.
In condensed matter physics, states with imaginary frequency arise frequently in disordered dissipative systems, in the form of purely relaxational overdamped modes, and even in heated crystals~\cite{Keyes_solid}. They play a crucial role in liquids and glasses where they are often called Instantaneous Normal Modes (INMs)~\cite{Keyes,Stratt,Nitzan} and correspond to saddle-points, with negative eigenvalues, in the energy landscape~\cite{broderix,debenedetti,sciortino}.

In all cases, defining the spectrum, or density of states (DOS), of systems that contain saddle points is challenging~\cite{parisi}. 
The major obstacle resides in the fact that the Plemelj identity, that formally provides the connection between the propagator and the DOS, is defined on the real axis, hence only states with real frequencies $\omega$, and positive eigenvalues $\lambda = \omega^{2}$ of the Hessian, are allowed. 
As a consequence, so far it has not been possible to analytically derive the DOS of systems with imaginary modes from a physical model, and the DOS of systems with INMs has been computed analytically only in $1d$ where negative eigenvalues are absent~\cite{parisiPRL}. 

In liquids (broadly defined to include plasma~\cite{murillo}), due to the presence of a large quantity of overdamped, exponentially-decaying modes with purely imaginary frequency (the INMs), this problem has hampered the derivation of a universal law for the DOS of vibrational excitations, $g(\omega)$. This is in stark contrast with the case of low-temperature crystals, where the vibrational modes are all real, and the Debye law $g(\omega) \sim \omega^{2}$ has served as a fundamental law for the DOS of solids since 1913. In solid glasses, the Debye law is still valid at the lowest frequencies, although hybridized with $\omega^{4}$ modes due to anharmonicity~\cite{Gurevich,Lerner}. 
In liquids, we know from numerical simulations~\cite{Stratt} that, as highlighted in  ~\cite{Rabani}, $g(\omega) \sim \omega$ at low frequency, but none of the theoretical models have been able to reproduce this law analytically.
This is due to the impossibility of including modes with negative eigenvalues which dominate the low-energy part of the spectrum.

Here we provide a solution to this longstanding problem, by developing the fundamental law for the DOS of liquids, which takes the imaginary modes into account analytically.
The key step in our derivation is the analytic continuation of the Plemelj identity to the complex plane. This leads to the possibility of defining a DOS for systems with imaginary frequency/energy. Our analytical model is successfully tested against numerical simulations from the literature on model (Lennard-Jones) liquids.

The vibrational density of states (DOS) of condensed matter systems is defined as
\begin{equation}
g(\omega)=\frac{1}{3\,\mathcal{N}}\sum_j\delta(\omega-\omega_{j})\label{DOSstart}
\end{equation}
with the index $j$ labelling different normal modes. Here $\mathcal{N}$ denotes the total number of atoms in the medium.
We consider decaying modes which are purely relaxational, as they arise e.g. from an overdamped Langevin dynamics in a liquid:
\begin{equation}
\frac{d\mathbf{v}}{dt}=-\Gamma \,\mathbf{v}, ~~~ \Gamma=1/\tau \label{lange}
\end{equation}
where $\mathbf{v}$ is the particle velocity, and $\tau$ is the relaxation time. The Langevin differential operator, associated to \eqref{lange}, is given by $ \mathfrak{L}\,:=\,d/dt\,+\,\Gamma$
and the corresponding Green's function is obtained by solving $\mathfrak{L}\,G(t,x)\,=\,\delta(t,x)$.
After Fourier transforming the previous identity, we obtain the following form for the Green's function:
\begin{equation}
G(\omega)=\frac{1}{i\,\omega + \Gamma}\,.\label{greenfinal}
\end{equation}

In order to relate the Green function for the single INM \eqref{greenfinal} with the total density of states \eqref{DOSstart}, we need to use the generalized form of Plemelj identity (see \cite{Julve,Julve2010}) valid for integration paths that do not necessarily lie on the real line,
\begin{equation}
\frac{1}{z-z'+i\,0^{+}}=P\left(\frac{1}{z-z'}\right)-i\,\pi\, \delta (z-z')
\label{Plemelj}
\end{equation}
where $z$ and $z'$ are arbitrary points in the complex plane region $\mathcal{D}^{+}$, which coincides with the complex plane $\mathcal{C}$ minus the wedge $-\pi/4 < \arg(z) < 5\pi/4$ and $P$ indicates the Cauchy principal value. 
This limitation arises upon considering the derivation of the Plemelj identity via distribution theory. The starting point of the derivation is usually given by an integral:
\begin{equation}
\int_{0}^{\infty} dx\, e^{-i z x}e^{-\epsilon x}, \nonumber
\end{equation}
where, clearly, as long as $z$ is real, no fundamental problem appears and the integral converges everywhere. 
Typically~\cite{Vladimirov}, one then proceeds by showing that this integral is equal to what one obtains upon applying the Fourier transform to the Heaviside function in $\mathcal S$ (the space of Schwartz distributions)~\cite{Schwartz}, which then leads to \eqref{Plemelj} above with $z,z'$ all real variables. 

Now, the analytic continuation of this integral, i.e. promoting $z$ to be a complex variable, leads to a divergence for $\text{Im} \,z < 0$. Following Zeldovich~\cite{Zeldovich}, the Gaussian regularization 
\begin{equation}
J(z,\lambda)=\int_{0}^{\infty}dx \,e^{-\lambda x^{2}}e^{i z x}
\label{Zeldovich}
\end{equation}
can be used (upon subsequently taking the limit $\lambda \to 0^{+}$). Following Ref.~\cite{Julve}, one can show that \eqref{Zeldovich} converges everywhere in the complex plane apart from the wedge defined by $-\pi/4 < \arg(z) < 5\pi/4$. 
Hence, one retrieves the generalized Plemelj identity \eqref{Plemelj} valid for arbitrary pathways in the allowed complex plane region $\mathcal{D}^{+}$.

We therefore write:
\begin{equation}
G(z)=\frac{1}{(\xi + i\,\omega) -(-\Gamma + \xi)+i\,0^{+}}
\end{equation}
thus identifying $z\equiv i\,\omega +\xi$, and $z'\equiv -\Gamma + \xi$, where $\xi$ is real. 
Next, we apply the generalized Plemelj identity (\eqref{Plemelj}) to $G(z)$ to evaluate the DOS
\begin{equation}
\frac{1}{(\xi + i\,\omega)  -(-\Gamma + \xi)+i\,0^{+}}=P\left(\frac{1}{z-z'}\right)-i\,\pi\, \delta (z-z')
\end{equation}
and, upon taking the imaginary part of the l.h.s. , we thus obtain:
\begin{align}
g(z)\equiv \delta(z-z')&=-\frac{1}{3\,\pi\,\mathcal{N}}\,\text{Im}\left[\frac{1}{i\omega  -(-\Gamma)+i\,0^{+}}\right]\\
&=\frac{1}{3\,\pi\,\mathcal{N}}\,\frac{\omega}{\omega^{2}+\Gamma^{2}}\,.\label{final}
\end{align}
The above \eqref{final} is our main result and provides a simple yet universal law for the VDOS of liquids. \eqref{final} is the density of state for a single INM with Green function \eqref{greenfinal}. Given the linearity of the problem, we can generalize the result to a set of $j$ INMs with different relaxation times as:
\begin{equation}
    g_{\text{total}}\,=\,\frac{1}{3\,\pi\,\mathcal{N}}\,\sum_j\,\frac{\omega}{\omega^2\,+\,\Gamma_j^2}\,,\label{total}
\end{equation}
where $\Gamma_j$ is the relaxation rate of the $j$-th INM.\\
Expanding \eqref{total} in the limit of low frequency, $\omega \ll \Gamma_j$, we obtain
\begin{equation}
g(\omega)=\alpha\, \omega\,+\,\mathcal{O}(\omega^3)\,,\quad \alpha\,=\,\sum_j\,\frac{1}{3\,\pi\,\mathcal{N}\,\Gamma_j^2}\,,\
\label{result}
\end{equation}
which recovers a common trend observed in many molecular simulation studies of liquids, where $\alpha$ is treated as a fitting parameter~\cite{Rabani,daligault}. 

\begin{figure}[ht]
    \centering
    \hspace{0.27cm} \includegraphics[width=\linewidth]{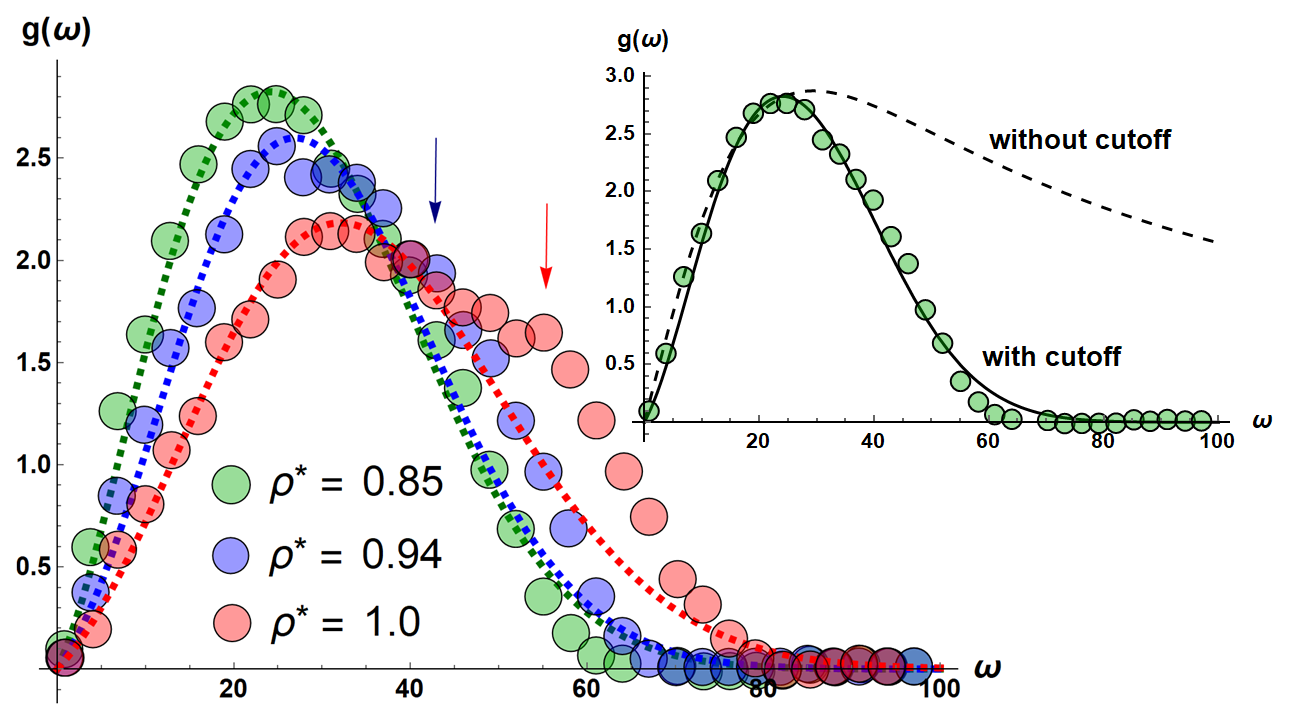}
    
    \vspace{0.4cm}
    
     \includegraphics[width=0.9 \linewidth]{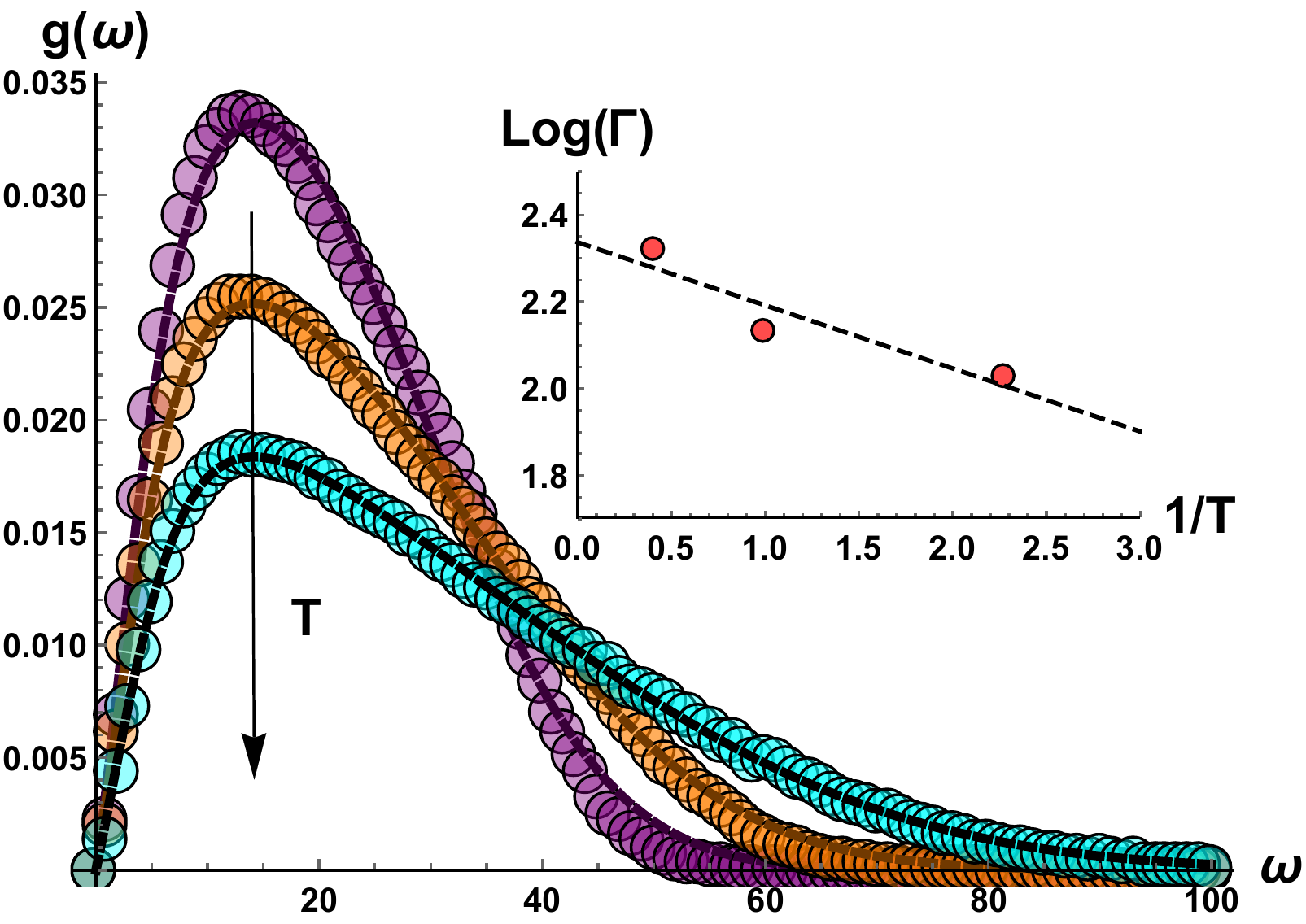}
     
    \caption{\textbf{Top Panel.} Data of the density of states taken from \cite{Rabani} and fitted with the analytic function \eqref{final}. A Gaussian (Debye) cutoff is added to take into account the large frequency fall-off. The arrows indicate the presence of possible relics of the van Hove singularities (which are not captured by our model) upon moving towards the solid phase. The original simulations were done using the standard LJ potential calibrated for Argon, and with $N$ in the range $108-256$. See \cite{Rabani} for more details of the simulation protocol. The inset shows the fit without the Gaussian cutoff.
    \textbf{Bottom Panel.} Data of the DOS of binary Kob-Andersen LJ liquids taken from \cite{Douglas} and fitted with the analytical formula \eqref{final}. A Gaussian (Debye) cutoff is added to take into account the large frequency fall-off. The inset shows that the fitted relaxation time $\Gamma$ in \eqref{result} as a function of temperature $T$ behaves according to the Arrhenius law, as expected for equilibrium liquids~\cite{Rabani}. The size in the original simulations data was set to $N=1000$ and the DOS was averaged over 100 independent realizations. Simulations were performed with a standard Nosé-Hoover  thermostat in the NVT ensemble.}
    \label{fig:unica}
\end{figure}

In \eqref{result}, the various relaxation rates sum up in parallel:
\begin{equation}
    \frac{1}{\Gamma_{\text{total}}^2}\,=\,\sum_j\,\frac{1}{\Gamma_j^2}\label{sum}
\end{equation}
implying that, in presence of a separation of scales $\Gamma^* \ll \Gamma_2,\Gamma_3,\dots$, the average relaxation rate $\Gamma$ would be given by the smallest of them, i.e. $\Gamma^*$. This is tantamount to saying that the low frequency dynamics is governed by the longest living imaginary mode -- a well-known result in the realm of hydrodynamics and effective field theory around equilibrium. In the rest of the paper, whenever we will discuss the relaxation rate, we will mean the total one defined in \eqref{sum}.\\

We now present predictions of the main result of our paper, \eqref{result}, in comparison with numerical simulations data of simple liquids, i.e. the Lennard-Jones (LJ) system. It is worth recalling that $g(\omega)$ is obtained numerically (from diagonalization of the Hessian matrix of instantaneous snapshots of particle positions) by retaining also the imaginary frequencies, because  $g(\omega) \equiv g(|\omega|)$~\cite{Keyes}. Hence, in the following, $\omega$ stands for the absolute value of the excitation frequency.

In Fig.\ref{fig:unica} we show the comparison between the model predictions and original MD simulations data of LJ liquids from Ref.~\cite{Rabani}. The model \eqref{result} uses an effective relaxation time $\Gamma$ as a fitting parameter, besides the normalization prefactor.

The simulation data can be nicely fitted using \eqref{result} with just two parameters up to the maximum of the DOS, after which a Gaussian cut-off $\sim e^{-(\omega/\omega_D)^{2}}$ is needed to capture the fall-off (note that in low-$T$ solid glasses a simple exponential cut-off is instead used~\cite{Chumakov}). In the dataset at the highest density, a small peak becomes visible, which is a relic of a pseudo-van Hove singularity upon approaching the solid state. 

In Fig.\ref{fig:unica}, we compare predictions of \eqref{result} with the more recent MD simulations data from Ref.~\cite{Douglas}, also for the LJ liquid. Also in this case, excellent agreement between \eqref{result} and simulations is found.
As a further confirmation of the validity of our Eq. \eqref{result}, we have also checked that the mean relaxation time $\Gamma$ follows the Arrhenius law as a function of $T$, as expected for equilibrium liquids~\cite{Rabani}, as shown in the inset of Fig.\ref{fig:unica}.
We found that the activation energy for these relaxations is $\approx 0.15 \epsilon$, where $\epsilon$ is the depth of the LJ attractive well, whereas the attempt frequency is $\sim 10 \epsilon/\hbar$, consistently about 10 times the escape rate from the LJ well. Since $\Gamma$ corresponds to the frequency of the maximum of the DOS (obtainable by setting the derivative of the summand in \eqref{total} equal to zero), this estimate gives an upper bound for the thermally-activated relaxation processes from anharmonic saddle points in the energy landscape.

This comparison shows that, especially in the low frequency region of the DOS up to the maximum, the spectrum is dominated by the relaxational modes or INMs. At higher frequencies, also phonons are present in the DOS~\cite{TrachenkoDOS}, which are not described by our minimal model, but can be included in future work. At present, the mean relaxation time $\Gamma$ acts as an effective parameter, which effectively takes into account also other vibrational excitations that are not explicitly implemented in the model and may become important around the frequency of the maximum and above.

In sum, the proposed theoretical model provides a universal law $g(\omega) \sim \omega$ for the low-frequency DOS of liquids, in agreement with observations from many simulations studies in the literature \cite{Rabani,Douglas,Keyes,Stratt}. This law plays for liquids the same pivotal role that the Debye law $g(\omega) \sim \omega^{2}$ plays for solids, and it explains e.g. that the liquids are mechanically unstable (by leading to diverging negative nonaffine contributions to the shear modulus~\cite{Palyulin}) and that INMs may play a role also for the thermal properties of solids. Furthermore, it could explain the observation of DOS scalings $\sim \omega^{\alpha}$ with $\alpha \in [1,2]$ in glasses~\cite{Suck} and other complex systems.
This law has not been derived before due to the difficulty of dealing with the imaginary frequencies (associated with saddle points in the energy landscape~\cite{broderix,debenedetti}) that dominate the low-frequency spectrum of liquids. In this work we have solved this problem by analytically continuing the Plemelj identity to the complex plane using input from recent  developments~\cite{Julve,Julve2010}.
This methodology is general and extends far beyond the case of liquids or glasses. Further applications of our result will lead to the possibility of analytically describing the DOS of unstable states in quantum mechanics and the spectrum of black holes where imaginary modes and quasinormal modes play an important role.

\section*{Acknowledgements}
A.Z. acknowledges financial support from US Army Research Office, contract nr. W911NF-19-2-0055. M.B. acknowledges the support of the Shanghai Municipal Science and Technology Major Project (Grant No.2019SHZDZX01) and
of the Spanish MINECO “Centro de Excelencia Severo Ochoa” Programme under grant
SEV-2012-0249.


\bibliography{amo}

\end{document}